\documentstyle[12pt,a4]{article}
\newcommand{\be}{\begin{displaymath}}
\newcommand{\ee}{\end{displaymath}}
\newcommand{\bea}{\begin{eqnarray}}
\newcommand{\eea}{\end{eqnarray}}
\newcommand{\ben}{\begin{equation}}
\newcommand{\een}{\end{equation}}
\newcommand{\kl}{{\rm K_L}}

\newcommand{\psiks}{{\rm J}/\psi\, {\rm K_S}}
\newcommand{\jpsi}{{\rm J}/\psi}
\newcommand{\pipic}{\pi^+ \pi^-}
\newcommand{\bb}{{\rm b \overline{b}}}
\newcommand{\ptr}{P_{\rm T}}
\newcommand{\bs}{{\rm B_s}}
\newcommand{\bsb}{{\rm \overline{B}{}_s}}
\newcommand{\bo}{{\rm B}^0}
\newcommand{\bob}{\overline{\rm B}{}^0}
\begin{document}
\title{\bf
\vspace{2.5cm}
Review of CP Violation Studies \\with B-Mesons at LHC
\vspace{10mm}}
\author{Tatsuya Nakada \vspace{5mm}\\
 Paul Scherrer Institute\\
CH-5232 Villigen-PSI, Switzerland \vspace{5mm}}
\date{19 December 1994}
\maketitle
\vspace{2cm}
%
\begin{abstract}
The Large Hadron Collider (LHC) proposed at CERN will be the ultimate source of
B-mesons. With the large number of B-mesons expected at LHC, a real precision
test
of CP violation in B-meson decays will become possible. There are already
several
efforts made to explore this possibility and a summary of those activities is
presented.
\end{abstract}
\vspace{2cm}
\begin{center}
Plenary talk given at \\
1994 International Workshop on B Physics \\
Physics Beyond the Standard Model at the B Factory\\
Nagoya, October 1994
\end{center}
\clearpage
\section{Introduction}
CP violation has been with us for more than 30 years \cite{cite-cronin} and its
origin still remains one of the unsolved problems in particle physics
\cite{cite-lee}. The standard model with three families of the left-handed
quark
doublet \cite{cite-cecilia} can explain observed CP violation in $\kl$ decays
through the complex Cabibbo-Kobayashi-Maskawa quark mass mixing
matrix \cite{cite-ckm}. However, it cannot be excluded for the moment that CP
violation is generated by a mechanism beyond the standard model
\cite{cite-wolfen}.

In the neutral kaon system, continuous experimental efforts have been made
to measure the isospin dependence of CP violation in the $\kl\rightarrow 2\pi$
decays predicted by the standard model. The expected signal is very small and
such a
dependence is not yet experimentally established \cite{cite-nakada}. Two new
experiments are in preparation.

Due to the complication introduced by the strong interaction at low energies,
theoretical predictions for CP violation cannot be made very accurately in the
kaon system. On the other hand, it is now generally accepted that such a
difficulty
is much smaller for particular decays in the B-meson system and the standard
model
can make accurate predictions once the four parameters of the CKM matrix become
known \cite{cite-bigi}. Therefore, it is worthwhile to measure the effects of
CP
violation in the B-meson system with an accuracy of a few \%. Another advantage
of
the B-meson is that it can decay into many different final states. The standard
model predicts a definite pattern of CP violation for different final states.
With
a large number of B-mesons, one can study such a pattern \cite{cite-wyler}. A
good
example is CP violation in $\bs$ decaying into $\jpsi \,\phi$. While the
standard
model expects negligible CP violation, a superweak model predicts a sizable
effect \cite{cite-gm}.

The standard model predictions for CP violation in the B-meson decays are
best illustrated by the unitarity triangle. Three sides of the
triangle are obtained from ${\rm b}\rightarrow {\rm c}+{\rm W}^-$ and
${\rm b}\rightarrow {\rm u}+{\rm W}^-$ decays and $\bo$-$\bob$
oscillations and will be well measured by CLEO and four LEP experiments.
The three angles, often referred as $\alpha$, $\beta$ and $\gamma$ can be
measured
from CP asymmetries in the B-meson decays. CP violation in the B-meson decays
could
be seen first by BaBar at SLAC, BELL at KEK, HERAB at DESY and possibly CDF
at FNAL \cite{cite-ball} before LHC.

The goal of the CP violation study at LHC is the ``precision'' test. Using a
large number of $\sim 10^{10}$ to $>10^{12}$ B-mesons expected at LHC, $\alpha$
and $\beta$ can be measured to an accuracy of $\le 0.01$. Measurements
of $\gamma$, of the mass difference between the two mass eigenstates of the
$\bs$-meson system through $\bs$-$\bsb$ oscillations and of the rare B-meson
decays
\cite{cite-ali} will be also unique contributions from LHC. In this article,
predicted experimental capabilities of proposed LHC detectors are summarised.

\section{General Consideration}
The b-quark cross sections are estimated to be $\sim 1~\mu$b in fixed target
mode
and $\sim 500~\mu$b in collider mode at LHC. $\bs$-mesons are also produced.
The
fractions of the b-quark production in the p-p interactions are
$\sim 0.0025\%$ in fixed target mode and $\sim 0.5\%$ in collider mode. The
large
b-quark cross section and $\sigma_{\rm b \overline{b}}/\sigma_{\rm total}$
ratio
are clear advantages for working with collider mode. It may be
noted that the current fixed target charm experiments operate with $\sigma_{\rm
c
\overline{c}}/\sigma_{\rm total}$ approximately $0.5\%$.

At LHC, the detector has to cope with an event rate of $40$~MHz with an  even
higher
interaction rate. The trigger, in particular the first-level trigger, must be
fast
and very selective. An experiment working in the fixed target mode can have
some
advantages in the trigger.

Due to the large b-quark mass, b-quark decays produce particles ($\mu$, e and
hadrons) with large momenta. In the p-p interaction, this translates into the
production of particles with a large transverse momentum ($\ptr$) respect to
the beam
axis. Muons can be identified fast and easily and the
number of muons in one event is small. Therefore, a large $\ptr$ muon trigger
is a
simple and effective first-level trigger. On the other hand, the trigger is
sensitive only to the semileptonic decays of the b-quark and to final states
with
$\jpsi$. This results in a low trigger efficiency.

The trigger efficiency will be roughly doubled if one can use the large $\ptr$
electron (and positron) in the first-level trigger. This is more difficult than
using muons due to many sources of background such as $\pi^0$ Dalitz decays and
photon conversions in the detector material. The small event multiplicity in
fixed
target mode could be an advantage in this respect \cite{nakada-2}.

Large $\ptr$ hadrons can be used effectively for the trigger in fixed target
mode.
It is well known that the average $\ptr$ of pions in normal p-p interaction
events increases with the centre of mass energy.  Therefore, the large $\ptr$
hadron trigger is not as effective in collider mode \cite{nakada-2}.

Another characteristic of b-quark decays is the displaced secondary vertex. In
fixed target mode, the average flight-length of a b-hadron is a few cm. If the
production target is point-like, the primary vertex is a priori known and a
very
fast trigger for selecting events containing tracks with a large impact
parameter
can be designed \cite{nakada-2}. In collider mode, the position of the primary
vertex is not known well due to the bunch length of the proton beam. Thus, the
primary vertex must be reconstructed first before selecting events containing
a track with a large impact parameter.

With a fixed target experiment using an extracted beam \cite{cite-waldi}, the
charged B-meson becomes ``visible'' by placing the vertex detector very closed
to
the production target. This could be useful for studying such decays as ${\rm
B}^+
\rightarrow \tau^+ \nu_{\tau}$ and $\bs \rightarrow \mu^+ \mu^-$ where the
background is a serious problem.

It is important to note that one can no longer
record all the events associated with the B-meson, and particular B-meson decay
modes of interest must be selected. This requires the online reconstruction of
events in the third-level trigger with all the detector information.

The ultimate limitation of an experiment may come from radiation
damage. It is very conceivable that LHC will produce more B-mesons than an
experiment can really use due to the radiation damage to the detector.

\section{General Purpose Detectors}
ATLAS \cite{atlas} and CMS \cite{cms} are two general purpose collider
detectors
designed to perform high $\ptr$ physics such as studies of the top quark and
search
for the Higgs and supersymmetric particles in the p-p interactions at LHC in
the
central region. The b-quark is an important tool for high $\ptr$ physics.

With the increasing interest in b-physics itself, the two collaborations
started to investigate the capabilities of their detectors to study b-physics
such
as CP violation in the B-meson decays. Those studies are also influencing the
design of the detectors, in particular the vertex detectors.

It is expected to take several years for LHC to reach its design luminosity
of ${\cal L}\approx 10^{34}$ ${\rm cm}^{-2}{\rm s}^{-1}$ which is required to
fully
exploit LHC for high $\ptr$ physics. Thus, b-physics will be an important
physics
programme for ATLAS and CMS during the first few years of the LHC operation.
Once
LHC achieves the design luminosity, b-physics will become exceedingly difficult
due
to the large background.

Both ATLAS and CMS have an excellent muon detection capability and the muon is
used in the first-level trigger. ATLAS uses a single muon with $\ptr \ge 6$
GeV.
For the B$\rightarrow \psiks$ decay mode, the trigger muon can be generated by
the
muons from the ${\rm J}/\psi$ decay or from the semileptonic decay of the
partner
b-quark which is used as the tag.

The CMS first-level trigger consists of a single muon with $\ptr \ge 10$
GeV or two muons with $\ptr \ge 3 \sim 5$ GeV. The double muon
trigger is very effective for B-meson decay final states with ${\rm J}/\psi$.
The single muon trigger is mainly sensitive to the semileptonic decay of the
partner b-quark used for the tag.

The excellent detection capability for the electron allows ATLAS to reconstruct
${\rm J}/\psi \rightarrow {\rm e}^+{\rm e}^-$. Due to the strong
magnetic field of the detector (4 T), CMS has difficulty to use the
electron channel.

\begin{table}[t]
\caption{Expected performances of the general purpose detectors at LHC for
$\sin
2\alpha$ and $\sin 2\beta$ using the time integrated method.}
\label{tab-general}
\begin{center}
\begin{tabular}{|l||c|c||c|c|}
\hline
\rule{0mm}{4.5mm} No.($\bb$)/$10^7$ sec & \multicolumn{4}{c|}{$5 \times
10^{12}$} \\
\hline  Measurement &\multicolumn{2}{c||}{$\sin \, 2\alpha$ from $\pipic$}
&\multicolumn{2}{c|}{$\sin \, 2\beta$ from $\psiks$}\\
\hline Experiment &ATLAS&CMS&ATLAS&CMS\\
\hline
\multicolumn{5}{|c|}{No.(reconstructed ``final state''+tag)}\\
\hline
$\bullet$ $\mu$-tag&  $3070$& $3400$ &\multicolumn{2}{c|}{${\rm J}/\psi
\rightarrow \mu^+ \mu^-$}\\
\cline{4-5} & & &$3847$&$9200$ \\
\cline{4-5} & & &\multicolumn{2}{c|}{${\rm J}/\psi
\rightarrow {\rm e}^+ {\rm e}^-$} \\
\cline{4-5} & & & $6041$&-\\
\hline
$\bullet$ e-tag &- &- &\multicolumn{2}{c|}{${\rm J}/\psi
\rightarrow \mu^+ \mu^-$}\\
\cline{4-5} & & & $4322$ & -\\
\hline total & $3070$ & $3400$& $14210$& $9200$\\
\hline background/signal & $1.67$ & $0.84$ & $\sim 0.1$ & $\sim 0.1$ \\ stat.
information & $0.71$ & $0.47$ & $0.62$ & $0.47$ \\
\hline
$^*\sigma_{\rm ststistical}$ & $0.08$ & $0.09$ & $0.028$ & $0.047$ \\
\hline
\multicolumn{5}{c}{\scriptsize $^*$Including the statistical fluctuation in the
background.}
\end{tabular}
\end{center}
\end{table}

Both experiments have improved their vertex detectors by placing their first
plane
much closer to the beam than the original designs shown in the letters of
intent.
The new designs provide a much better impact parameter resolution which reduces
the
background in the reconstructed B-mesons and improves the eigentime resolution
of
the B-meson. However, the radiation damage while operating at the nominal LHC
luminosity  becomes a serious concern.

Table \ref{tab-general} summarises the expected performance for measuring $\sin
\,
2\alpha$ and $\sin\, 2 \beta$ by ATLAS and CMS using the CP asymmetries
obtained
from the decay time integrated rates. It is assumed that LHC will run with an
average luminosity of
$10^{33}$ for
$10^7$ s, i.e. roughly one year. The quoted errors are only statistical. It
shows
that $\sin 2\beta$ can be measured very well. For the measurement of $\sin
2\alpha$,
the large amount of remaining background is a worry. The background comes
mainly
from other two-body decay modes of b-hadrons such as ${\rm B\rightarrow K}\pi$
and
${\rm B_s \rightarrow KK}$. The momentum resolution is not sufficient to
distinguish them from the B$\rightarrow \pipic$ decay.

One way to separate the background in ATLAS and CMS, which have no special kaon
and
pion identification capabilities, is to study the decay time distribution. The
background events are expected to decay (almost) purely exponentially. With the
absence of the penguin diagram, the decay time distribution for $\pipic$ events
is
given by
\begin{displaymath}
 e^{-\, {\it\Gamma} \, t} \left( 1\pm \sin\, 2\beta \times \sin \,\Delta m \,
t
\right)
\end{displaymath}
where $\Delta m $ and ${\it \Gamma}$ are the mass difference between the
two weak eigenstates and the decay width of the neutral B-mesons, respectively.
Another advantage of studying the decay time distribution is that the errors on
both $\sin 2\alpha$ and $\sin 2\beta$ can be reduced by $\sim 20 \%$ due to
the increased statistical sensitivity of the method \cite{cms2}.

\section{Dedicated Detectors}
\subsection{Past}
Three different approaches to perform B-physics in a dedicated way, COBEX
\cite{cobex}, GAJET \cite{nakada-2} and LHB \cite{cite-waldi}, were initiated.
COBEX
proposed to work in collider mode and GAJET and LHB in fixed target mode. An
internal gas-jet target was considered for GAJET while LHB considered
extracting the
halo of the LHC beam parasitically using a bent crystal. All three experiments
were
designed to run for many years in different luminosity conditions of LHC.

All three detectors were forward spectrometers equipped with a Si vertex
detector
very close to (or in) the beam, large aperture magnet(s) with a tracking
system,
a particle identification system capable of the $\pi$/K/p separation over all
the
necessary kinematic range, electromagnetic (and hadronic for GAJET and LHB)
calorimeter(s) and a muon system.

Compared with the two fixed target experiments, COBEX benefited from the larger
b-quark production cross section in collider mode. GAJET emphasised its simple
and
effective impact parameter trigger strategy based on the point-like target
geometry
combined with the large $\ptr$ lepton and hadron triggers. LHB deployed a
vertex
detector system close to the production target where most of the B-mesons
decay.

General advantages of a dedicated detector compared to a general purpose
detector
are the following:
\begin{itemize}
\item
The forward spectrometer geometry allows a more efficient muon $\ptr$ trigger
with a
lower threshold value.
\item
The vertex detector system close to the beam provides a better vertex
resolution. This is important in particular for studying $\bs$-$\bsb$
oscillations
and CP violation in $\bs$ decays.
\item
The particle identification system reduces the background in the B$\rightarrow
\pipic$ decay mode generated by other two-body decay modes of b-hadrons to a
negligible amount. It also reduces the combinatorial background and the
many-body decay modes of B- and $\bs$-mesons can be reconstructed. This
allows measurements of the third angle of the
unitary triangle, $\gamma$, CP asymmetries expected to be very small in the
standard model and CP asymmetries in charged B-meson decays.
\end{itemize}

\begin{table}
\caption{Expected performances for past proposed dedicated experiments at LHC.}
\label{tab-dedicated}
\begin{center}
\begin{tabular}{|l||c|c|c|}
\hline
Experiment & COBEX & GAJET & LHB \\ \hline
\rule{0mm}{4.5mm}
No.($\bb$)/$10^7$ sec & $4 \times 10^{12}$ & $2 \times 10^{10}$ & $7.7 \times
10^{9}$\\
\hline
\multicolumn{4}{|c|}{First-level trigger} \\
\hline
High $\ptr$ & $\mu$ & $\mu$, e, hadron &$\mu$, e \\
Large impact parameter & (only at low $\cal L$) & yes & - \\
\hline
\rule{0mm}{4.5mm}
Tagging method & $\mu$ (${\rm K}^\pm$) &  $\mu$, e, ${\rm K}^\pm$ &
$\mu$, e, ${\rm K}^\pm$, ${\rm B}^\pm$ \\
\hline
\multicolumn{4}{|c|}{$\sin 2\alpha$ from $\pipic$}\\
\hline
No.(reconstructed $\pipic$+tag) &$30000$ &$4500$ & $3200$ \\
background/signal & $<0.16$ & $0.3$ & $<0.1$ \\
\hline
$\sigma_{\sin 2\alpha}$ statistical &
$0.015$ & $0.04$ & $0.07$ \\
\hline
\multicolumn{4}{|c|}{$\sin 2\beta$ from $\psiks$}\\
\hline No.(reconstructed $\psiks$+tag) &$270000$ &$10350$ & $13000$ \\
\hline
$\sigma_{\sin 2\beta}$ statistical &
$0.007$ & $0.03$ & $0.02$ \\
\hline
\multicolumn{4}{|c|}{final states useful to measure $\gamma$}\\
\hline
\rule{0mm}{4.5mm}
$\bs \rightarrow {\rm D_s \, K}$/branching ratio& $^*1.8 \times 10^8$ & $^*8.3
\times 10^7$ &
$^*2.8
\times 10^7$ \\
\rule{0mm}{4.5mm}
${\rm B}^+\rightarrow {\overline{\rm D}^0 \, {\rm K}^+}$/branching ratio & - &
- &
$^*1.2\times 10^8$
\\
\rule{0mm}{4.5mm}
${\rm B}^0 \rightarrow \overline{\rm D}^0 \, {\rm K}^{*0}$/branching ratio & -
&
$^*1.9 \times 10^8$ & - \\
\hline
\multicolumn{4}{c}{\scriptsize $^*$Relevant branching fractions must be
multiplied to obtain the actual number of reconstructed events. }
\end{tabular}
\end{center}
\end{table}

Table \ref{tab-dedicated} summarises the performances of the three dedicated
experiments. Dedicated experiments indeed do much better in the difficult decay
mode B$\rightarrow \pipic$ than general purpose experiments not only
statistically
but also in the reduction of the background.

\subsection{Current Status and Future}
Although the LHC committee (LHCC) has repeatedly confirmed the necessity of a
dedicated B-physics experiment as one of the baseline LHC experiments along
with
ATLAS, CMS and ALICE, none of the above three experiments was recommended
for submitting a technical proposal. Instead, LHCC requested the submission of
a
new letter of intent by a joint collaboration based on the collider mode with a
newly designed forward spectrometer with a vertex detector system placed in the
Roman pot \cite{schlein}. A collaboration containing most of the members of the
original three groups and many other institutes was formed to do this task. The
collaboration is in the process of optimising the detector and the trigger
strategy
and intends to submit the letter of intent by the end of February 1995. A
performance even better and more solid than that claimed by the originally
proposed
dedicated experiments is expected.

\section{Conclusions}
There are already active efforts to plan to measure CP violation in B-meson
decays at LHC. The two general purpose experiments, ATLAS and CMS will study
CP violation during the initial period of LHC running with less luminosity
than the design one. They can measure $\sin \, 2\beta$ using the B$\rightarrow
\psiks$ decay mode well and contribute to the $\sin \, 2\alpha$ measurement.
They
have an excellent mass resolution and a good B-meson decay vertex resolution.
Their
limitation is in the particle identification.

A dedicated B-physics detector at LHC will tackle the problem of CP violation
in
the B-meson decay for many years and try to measure the angles of the
unitarity triangle with a precision of $\stackrel{<}{\scriptstyle \sim} 0.01$.
It
will have a more efficient trigger for the b-quark events than the general
purpose
detectors. The capability of identifying pions, kaons and protons will ensure
clean
reconstruction of many different B-meson decay modes which is important to
study CP
violation in a complete way.

\section*{Acknowledgements}
The author thanks to the organizers for their hospitality during this
stimulating
conference. Discussions with many physicists working at ATLAS and CMS, in
particular P.~Eerola (ATLAS), D.~Denegri, R.~Horisberger and A.~Rubbia (CMS),
are
acknowledged. The author would like to thank the colleagues from the COBEX and
LHB
collaborations for useful discussions. Particular thanks are to the member of
the
GAJET collaboration for the stimulating work during the preparation of the
letter
of intent, in particular A.~Buijs, L.~L.~Camilleri, P.~D.~Dauncey, M.~Dracos,
P.~Galumian, Y.~Lemoigne, S.~Loucatos, J.-P.~Perroud and W.~Ruckstuhl. The
author
acknowledges Q.~Ingram for reading this manuscript.


\begin{thebibliography}{99}
\bibitem{cite-cronin}
J.~H.~Christenson et al., Phys. Rev. Lett. {\bf 13} (1964) 138.
\bibitem{cite-lee}
T.~D.~Lee, contribution to these proceedings.
\bibitem{cite-cecilia}
CP violation within the framework of the standard model is summarised by
C.~Jarlskog in these proceedings.
\bibitem{cite-ckm}
N.~Cabibbo, Phys. Rev. Lett. {\bf 10} (1963) 531.\\
M. Kobayashi and K. Maskawa, Prog. Theor. Phys. {\bf 49} (1973) 652.
\bibitem{cite-wolfen}
For a review on CP violations and physics beyond the standard model, see the
contribution from L.~Wolfenstein in these proceedings.
\bibitem{cite-nakada}
For a review on CP violation experiment in the kaon system, see for example\\
T.~Nakada, AIP Conference Proceedings 302, Lepton Photon Interactions, XVI
International Symposium (ed. P.~Drell and D.~Rubin, Ithaca, August 1993) p.425.
\bibitem{cite-bigi}This point is nicely summarised by\\
I.~I.~Bigi, Proceedings of the Second International Workshop on B-Physics at
Hadron
Machines (ed. P.~E.~Schlein, Le Mont-Saint-Michel, April 1994), Nucl. Inst.
and Meth. {\bf A351} (1994) 240.
\bibitem{cite-wyler}The pattern of CP violation in B-meson decays is summarised
by\\
D.~Wyler, Proceedings of the Second International Workshop on B-Physics at
Hadron
Machines (ed. P.~E.~Schlein, Le Mont-Saint-Michel, April 1994), Nucl. Inst.
and Meth. {\bf A351} (1994) 8.\\
See also contributions by I.~I.~Bigi and M.~Gronau in these proceedings.
\bibitem{cite-gm}J.-M.~G\'erard and T.~Nakada, Phys. Lett. B {\bf 261} (1991)
474.
\bibitem{cite-ball}See contributions from P.~Oddone (BaBar), S.~Suzuki (BELL),
A.~S.~Schwarz (HERAB) and N.~Lockyer (CDF) in these proceedings.
\bibitem{cite-ali}A.~Ali, contribution to these proceedings.
\bibitem{nakada-2}P.~D.~Dauncey et al. (GAJET Collaboration),
Proceedings of the Second International Workshop on B-Physics at Hadron
Machines (ed. P. E. Schlein, Le Mont-Saint-Michel, April 1994), Nucl. Inst.
and Meth. {\bf A351}  (1994) 147.
\bibitem{cite-feroni}F. Ferroni (RD22),
Proceedings of the Second International Workshop on B-Physics at Hadron
Machines (ed. P. E. Schlein, Le Mont-Saint-Michel, April 1994), Nucl. Inst.
and Meth. {\bf A351}  (1994) 183.
\bibitem{cite-waldi}R.~Waldi (LHB Collaboration),
Proceedings of the Second International Workshop on B-Physics at Hadron
Machines (ed. P. E. Schlein, Le Mont-Saint-Michel, April 1994), Nucl. Inst.
and Meth. {\bf A351}  (1994) 161.
\bibitem{atlas}P.~Eerola et al. (ATLAS),
Proceedings of the Second International Workshop on B-Physics at Hadron
Machines (ed. P. E. Schlein, Le Mont-Saint-Michel, April 1994), Nucl. Inst.
and Meth. {\bf A351}  (1994) 84.
\bibitem{cms}D.~Denegri et al. (CMS),
Proceedings of the Second International Workshop on B-Physics at Hadron
Machines (ed. P. E. Schlein, Le Mont-Saint-Michel, April 1994), Nucl. Inst.
and Meth. {\bf A351}  (1994) 95.
\bibitem{cms2} See for example A.~Kharchilava and D.~F.~Vit\`e (CMS).
CMS Technical Note TN/94-193 (1994)
\bibitem{cobex}S.~Erhan et al. (COBEX),
Proceedings of the Second International Workshop on B-Physics at Hadron
Machines (ed. P. E. Schlein, Le Mont-Saint-Michel, April 1994), Nucl. Inst.
and Meth. {\bf A351}  (1994) 132.
\bibitem{schlein}J.~Ellett et al., Nucl. Inst. and Meth. {\bf A317}  (1992) 28.
\end{thebibliography}
\end{document}